\documentclass[draft,grl]{AGUTeX}

\usepackage{lineno}
\linenumbers*[1]

  \usepackage[dvips]{graphicx}
 \setkeys{Gin}{draft=false}
\authorrunninghead{MILOCH AND VLADIMIROV}

\titlerunninghead{CHARGING OF SPINNING OBJECTS}

\authoraddr{W. J. Miloch,
Institute of Theoretical Astrophysics, University of Oslo, Box 1029 Blindern, N-0315 Oslo, Norway (w.j.miloch@astro.uio.no)}
\authoraddr{S. V. Vladimirov, School of Physics, The University of Sydney, Sydney, NSW 2006, Australia}

\begin{document}

%
%

\title{Charging of spinning insulating objects by plasma and photo\-emission}
%

%
%


\author{W. J. Miloch}
\affil{
Institute of Theoretical Astrophysics, University of Oslo, Box 1029 Blindern, N-0315 Oslo, Norway}
\affil{School of Physics, The University of Sydney, Sydney, NSW 2006, Australia}

\author{S.~V. Vladimirov}
\affil{School of Physics, The University of Sydney, Sydney, NSW 2006, Australia}



%
%
%

%
%


\begin{abstract}
The charging of spinning insulating objects by plasma and photo\-emission is studied with the particle-in-cell method. Unidirectional photon flux, different angular velocities of the object, and different plasma flow speeds are considered. Photo\-emission can lead to a positive total charge and electric dipole moment on the object. The spinning of the object redistributes the surface charge. The total object charge oscillates in time with the period matching the period of the object rotation. Plasma potential and density in the vicinity of the object oscillate with the same frequency. The plasma is rarefied close to a positively charged object, and the density wake diminishes when the object is charged negatively. The time averaged charge depends on the angular velocity of the object. 

\end{abstract}

%
%

%

\begin{article}

%
%

\section{Introduction}
The charging of an object by plasma is one of the basic problems in space and plasma physics. If only plasma currents are considered, the charge on the object is usually negative \citep{Svenes_Troim_1994, Vladimirov_Ostrikov_2005, Miloch_Pecseli_Trulsen_2007}.
The plasma flow introduces asymmetry to the object's charging and gives rise to wakes in the plasma density and potential \citep{Vladimirov_Nambu_1995, Melandso_Goree_1995, Vladimirov_Ishihara_1996, Ishihara_Vladimirov_1997}. This asymmetry is more pronounced for insulating than for conducting objects \citep{Miloch_Pecseli_Trulsen_2007, Miloch_Pecseli_Trulsen_2008}.
Photo\-electric current can lead to positively charged objects \citep{Shukla_Mamun_2002, Vladimirov_Ostrikov_2005}. If photo\-emission is due to a directed photon flux, the electric dipole moment can develop on insulating objects. There is a significant difference between the wake of the negatively and positively charged object, with a strong density rarefaction for the latter \citep{Miloch_Vladimirov_2008}.

The understanding of the charging of an object in a complex environment with sunlight and plasma flow is of concern for the operation of spacecrafts or sounding rockets \citep{Svenes_Troim_1994, Roussel_Berthelier_2004}. The charging of insulating components of such objects can result in strong potential differences between the shadow and sunlit sides. Changes of the plasma parameters in the vicinity of the object need to be accounted for when analyzing the instrument data \citep{Lai_Cohen_1986}. The problem can be further complicated by the spinning of an object, whether it is an intrinsic rotation of an asteroid or dust grain, or imposed for the attitude and stability control of a spacecraft or rocket \citep{Lee_Sun_2001, Kurihara_Oyama_2006}.

Analytical models for satellites spinning in sunlight demonstrated the development of potential barriers that decelerate photoelectrons and allow the sunlit side of the spacecraft to be negatively charged  \citep{Tautz_Lai_2006, Tautz_Lai_2007}. These models assume the satellite to be much smaller than the Debye length, and replace the Poisson equation with the Laplace equation for vacuum. Moreover, they do not account for the plasma flow.  If the plasma flow and intermediate sizes of the satellite with respect to the Debye length are considered, the problem becomes highly nonlinear. A self-consistent analytical model for such a problem is difficult to develop, and the numerical analysis seems appropriate. 

In this letter we discuss results from the particle-in-cell (PIC) simulations of spinning insulating objects in flowing plasmas exposed to unidirectional photon flux. The analysis is relevant for such objects as  satellites, rockets or asteroids in space, boulders on lunar surface, or dust grains in experimental devices \citep{Horanyi_1996, Fortov_Nefedov_1998, Khrapak_Nefedov_1999}. 

\section{Numerical code}
The analysis is carried out in two spatial dimensions in Cartesian coordinated using the PIC numerical code described in detail by \citet{Miloch_Pecseli_Trulsen_2007, Miloch_Pecseli_Trulsen_2008, Miloch_Vladimirov_2008, Miloch_Vladimirov_2008b}. The electrons and ions are treated as individual particles, with the ion to electron mass ratio being $m_i/m_e=120$. The plasma density is $n=10^{10}~\mathrm{m^{-2}}$. The collision\-less plasma flows in the positive $x$ direction, and three plasma drift velocities are considered $v_d=\{0, 0.5, 1.5\} C_s$, with $C_s$ being the speed of sound for adiabatic ions and isothermal electrons: $C_s=\sqrt{\kappa (T_e+\gamma T_i)/M}$, where $\gamma=5/3$. The electron to ion temperature ratios are $\zeta=T_e/T_i=\{5,100\}$, where $T_e=0.18$ $\mathrm{eV}$. 

A circular object of radius $r=0.375$ in units of the electron Debye length $\lambda_{De}$ is placed within a simulation box of size of $50 \times 50$ $\lambda_{De}$. Such an object  can be understood as an intersection of a cylindrical object in a three dimensional system, and it is initially charged only by the collection of electrons and ions. To represent a perfect insulator, each plasma particle that hits the surface remains at this position for all later times contributing to the local charge density. The object spins throughout the whole simulation with angular velocity $\omega$ of $0.5 \pi$, $2 \pi$ or alternatively $3 \pi$ in units of $\mathrm{rad}/\tau_i$, where  $\tau_i$ is the ion plasma period. A directed photon flux is switched on at $39\tau_i$, when it can be assumed that the surface charge distribution has reached a stationary level. The code is run for approximately $50\tau_i$. 

\begin{figure}
\noindent\includegraphics[width=20pc]{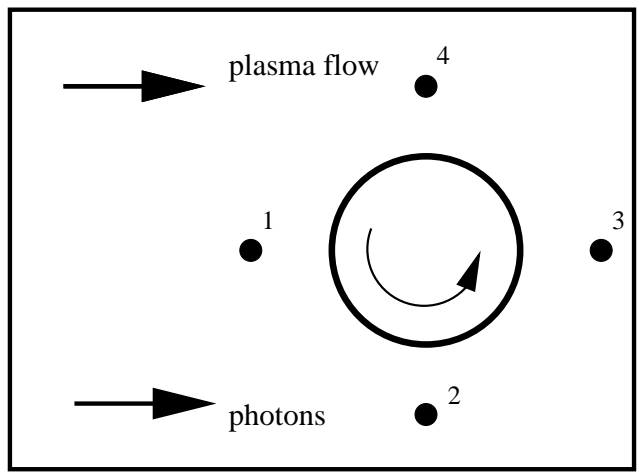}
\caption{Scheme of a typical numerical arrangement. Points labeled with numbers correspond to the probes for potential variations in the vicinity of the object. The object rotates anticlockwise. }
\label{fig:scheme}
\end{figure}

The code allows for an arbitrary angle of the photon incidence on the object. When a photon hits the object surface, a photo\-electron of energy 1 $\mathrm{eV}$ is produced at distance $l=sv\Delta t$ from the surface, where $s$ is an uniform random number $s \in (0,1]$,  $\Delta t$ is the computational time step, and $v$ is the photoelectron speed. Photoelectron velocity vectors are uniformly distributed over an angle of $\pi$ and directed away from the surface which is in accordance with Lambert's law. In this study the incoming photons are usually aligned with the direction of the plasma flow, i.e., $\alpha=0^{\circ}$.  As a control case, we also consider the angle of $90^{\circ}$ between the incoming ions  and the plasma flow ($\alpha=90^{\circ}$). The photon flux is $\Psi=1.25 \times 10^{19} \mathrm{~m^{-2}s^{-1}}$. The scheme of a typical numerical arrangement is shown in Fig.~\ref{fig:scheme}. Points close to the object's surface, that are labeled with numbers, indicate probes for potential variations that are described in more detail in Results section.

\section{Results}

\begin{figure}
\noindent\includegraphics[width=20pc]{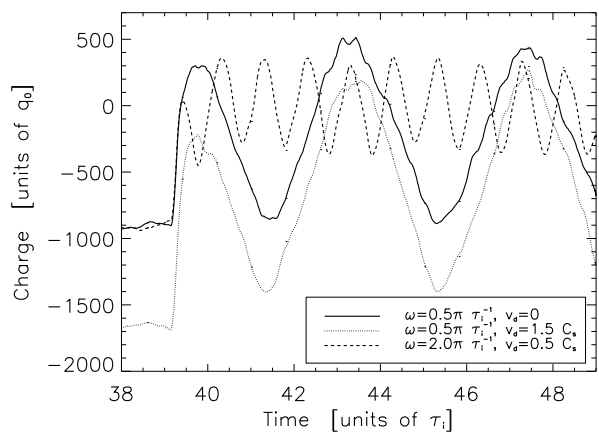}
\caption{The charging of spinning insulating object during photo\-emission for different angular velocities $\omega$ and speeds of the plasma flow $v_d$ for $\zeta=100$. The results are smoothed with a moving box filter for presentation.}
\label{fig:q}
\end{figure}

With the onset of radiation, the total charge on a spinning object becomes more positive and starts to oscillate in time. The period of oscillations matches the period of the full rotation of the object, see Fig.~\ref{fig:q}. The charge variations are large for slowly spinning objects. The mean charge depends on the plasma flow velocity. It reaches lower values for faster plasma flows $v_d$, and is more positive with higher angular velocities of the object $\omega$. The mean charge values for different $v_d$ and $\omega$ are summarized in Table \ref{tab:meanq}. Since there is little difference in the charging characteristic for $\zeta=5$ and $\zeta=100$, only the results for $\zeta=100$ are shown in Fig.~\ref{fig:q} and Table~\ref{tab:meanq}.

Photo\-emission due to unidirectional photons leads to the development of an electric dipole moment on the object. The electric dipole moment is initially antiparallel to the photon direction, and it co-rotates with the object. It vanishes at the certain angle between the incoming photon direction and the dipole moment, which increases with increasing $\omega$. The electric dipole moment antiparallel to the photon direction reappears after the full rotation of the object.

\begin{figure}
\noindent\includegraphics[width=20pc]{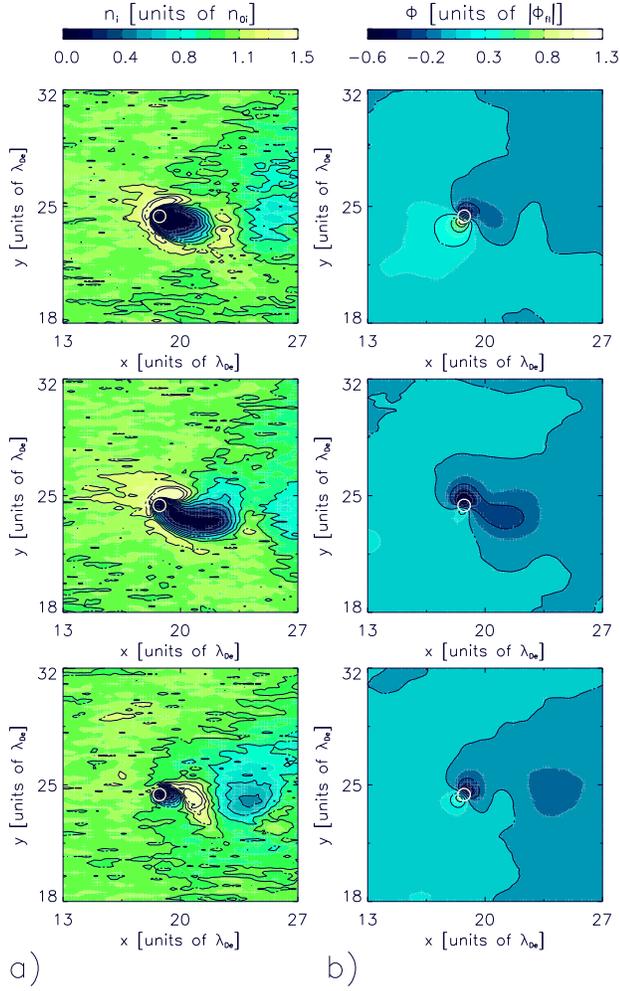}
\caption{Ion density (a) and electric potential (b) for a spinning insulating object during photo\-emission  with $\omega=2\pi ~\tau_i^{-1}$, $v_d=1.5C_s$, and $\zeta=100$. Different time instances are shown for which the total charge on the object is positive (top), the erosion of the wake begins (middle), and the total charge on the object is negative (bottom). }
\label{fig:wake}
\end{figure}

In flowing plasmas, the wake in the plasma density forms behind a spinning object. This wake oscillates in time. The region of rarefied plasma density has large spatial extent when the object is positively charged, while for a negatively charged object, the plasma density in the wake can be  enhanced and the ion focusing observed. The oscillatory nature of the wake is demonstrated in the ion density plots in Fig.~\ref{fig:wake}a). The transition between the ion wake and the enhanced ion density is asymmetric. The edge of the wake is distorted by the local enhancement in the ion density. This distortion grows in time and reduces the wake size. 

Similar rarefication in the ion density and scenario for the wake closure are observed for stationary plasmas, with the difference that the ion density is rarefied close to the positively charged side of the object and the wake does not form.

\begin{figure}
\noindent\includegraphics[width=20pc]{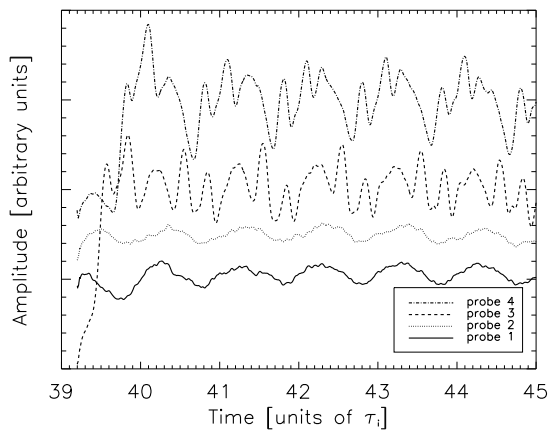}
\caption{Potential variations for different probes at distances $d=\lambda_{De}$ to the surface of the spinning object for $v_d=1.5C_s$, $\omega=2\pi ~\tau_i^{-1}$, and $\zeta=100$. The orientation of the probes with respect to the plasma flow and direction of photons is shown in Fig.~\ref{fig:scheme}. }
\label{fig:probes}
\end{figure}

Without photo\-emission, a symmetrical sheath forms around the object, while for flowing plasmas the wake behind the object is observed. The potential distribution in the vicinity of the object is governed by the photo\-emission and associated electric dipole moment. The oscillations of the total charge alter the potential distribution, and for flowing plasmas, the wake in the potential is highly distorted, see Fig.~\ref{fig:wake}b).
Since many instruments on the rocket and satellite payloads are installed on extended booms, it is vital to examine to what extend a spinning insulating body modifies the surrounding plasma. In Fig. \ref{fig:probes} we show potential variations at four point-like probes at distances $d=\lambda_{De}$ from the object surface in the direction parallel and perpendicular to the photon and plasma flows. The orientation of each of the probe relative to the plasma flow is shown in Fig.~\ref{fig:scheme}. For all probes, we observe oscillations with the periods equal to the rotation period of the object. The amplitudes of oscillations on the side of the object that is charged predominantly positive (i.e., probes 1 and 2) are moderate as compared to probes 3 and 4. For the latter probes, other strong, repeatable components are present in the signal. The amplitude of potential variations for probes 3 and 4 are twice as large as for other probes.

For temperature ratio $\zeta=5$, the total charge on the object is less negative than for $\zeta=100$ due to the ion mobility. The charging characteristic with photo\-emission is similar for both cases, with the negative part of oscillations being more negative for colder ions. The ion wake is smaller and less pronounced for warmer ions, but the analyses of potential oscillations close to the object are similar for both temperature ratios.

For the control case of the photons incidence angle of $90^{\circ}$ with respect to the plasma flow, the wake is spatially different from the case for  $\alpha=0^{\circ}$. However, the principal mechanism and oscillations of the total charge in time are the same. The mean charge is slightly higher for  $\alpha=90^{\circ}$ than for  $\alpha=0^{\circ}$.

\section{Discussion and Conclusions}
With the onset of the photo\-emission due to unidirectional photon flux, an electric dipole moment develops on the object. Most of the positive charge is localized on the illuminated side of the object, while the opposite side is predominantly negatively charged. For a spinning object, the photo\-electric current neutralizes negatively charged regions, but the photo\-emission rate can often be too low to charge them positively. Only after approximately one rotation period of the object, photo\-emission can lead to the positive total charge on the object and recovery of the electric dipole moment. This suggests that the initial charging of the object by photo\-emission is robust, and that the charge redistribution on the object's surface is insufficient to compensate for the residual charge with the spinning rates considered in this study.

This picture should be complemented by the ion and electron dynamics close to the object. In case of the plasma flow, the wake is formed. For spinning objects, this wake is distorted on one side. Spinning, negatively charged side of the object accelerates ions towards the wake and enhances the plasma density locally. The wake is further distorted due to the enhancement in the plasma density and neutralization of the positively charged regions on the surface. 

A spinning insulating object excites waves in the system with the wave frequency matching the frequency of the object rotation. While these waves are damped further away from the object, they are conspicuous close to the object's surface. Since the plasma density is inhomogeneous and the wake oscillates, the wave propagation is complicated. On the side predominantly positively charged, the variations in the potential distribution are smooth with a single dominant frequency observed. On the negatively charged side, where  the wake oscillations modify the plasma to a high degree, the potential variations are larger, and modulations of the main wave are observed.

Previous analytical works by \citet{Tautz_Lai_2006, Tautz_Lai_2007} did not include plasma dynamics around the object, and hence stationary solutions could be obtained. In the limiting case of a small object with respect to $\lambda_{De}$, one can expect such results with the conspicuous potential barrier for photo\-electrons, especially for fast spinning insulating objects. We have shown, however, that in the regions where the plasma dynamics around the object is important, the result will be different, and the photo\-emission and spinning of the object break the symmetry of the object charging even without the plasma flow. In particular, the satellite or rocket instrument readings can be influenced by the rotation, which has to be accounted for in data analysis. For a spinning satellite traveling through regions with different plasma densities and temperatures it will often be necessary to employ these two different models.


%
%
%
%
%
%

%
%
%
%

\begin{acknowledgments}
This work was in part supported by the Norwegian Research Council, NFR, and by the Australian Research Council, ARC.
\end{acknowledgments}

%
%

\begin{thebibliography}{20}
\providecommand{\natexlab}[1]{#1}
\expandafter\ifx\csname urlstyle\endcsname\relax
  \providecommand{\doi}[1]{doi:\discretionary{}{}{}#1}\else
  \providecommand{\doi}{doi:\discretionary{}{}{}\begingroup
  \urlstyle{rm}\Url}\fi

\bibitem[{\textit{Fortov et~al.}(1998)}]{Fortov_Nefedov_1998}
Fortov, V.~E., et~al. (1998), Dusty plasma induced by solar radiation under
  microgravitational conditions: an experiment on board the MIR orbiting space
  station, \textit{J. Exp. Theoret. Physics}, \textit{87}(6), 1087--1097.

\bibitem[{\textit{Hor\'{a}nyi}(1996)}]{Horanyi_1996}
Hor\'{a}nyi, M. (1996), Charged dust dynamics in the solar system,
  \textit{Annu. Rev. Astron. Astrophys.}, \textit{34}, 383--418.

\bibitem[{\textit{Ishihara and Vladimirov}(1997)}]{Ishihara_Vladimirov_1997}
Ishihara, O., and S.~V. Vladimirov (1997), Wake potential of a dust grain in a
  plasma with ion flow, \textit{Phys. Plasmas}, \textit{4}(1), 69--74.

\bibitem[{\textit{Khrapak et~al.}(1999)\textit{Khrapak, Nefedov, Petrov, and
  Vaulina}}]{Khrapak_Nefedov_1999}
Khrapak, S.~A., A.~P. Nefedov, O.~F. Petrov, and O.~S. Vaulina (1999),
  Dynamical properties of random charge fluctuations in a dusty plasma with
  different charging mechanisms, \textit{Phys. Rev. E}, \textit{59}, 6017.

\bibitem[{\textit{Kurihara et~al.}(2006)\textit{Kurihara, Oyama, Iwagami, and
  Takahashi}}]{Kurihara_Oyama_2006}
Kurihara, J., K.~I. Oyama, N.~Iwagami, and T.~Takahashi (2006), Numerical
  simulation of {3-D} flow around sounding rocket in the lower atmosphere,
  \textit{Ann. Geophys.}, \textit{24}, 89--95.

\bibitem[{\textit{Lai et~al.}(1986)\textit{Lai, Cohen, and
  McNeil}}]{Lai_Cohen_1986}
Lai, S.~T., T.~L. Cohen, H. A. amd~Aggson, and W.~J. McNeil (1986), Boom
  potential of a rotating satellite in sunlight, \textit{J. Geophys. Res.},
  \textit{91}(A11), 12,137--12,141.

\bibitem[{\textit{Lee et~al.}(2001)\textit{Lee, Sun, Tahk, and
  Lee}}]{Lee_Sun_2001}
Lee, H.~I., B.~C. Sun, M.~J. Tahk, and H.~Lee (2001), Control design of
  spinning rockets based on co-evolutionary optimization, \textit{Control
  Engineering Practice}, \textit{9}, 149--157.

\bibitem[{\textit{Melands{\o} and Goree}(1995)}]{Melandso_Goree_1995}
Melands{\o}, F., and J.~Goree (1995), Polarized supersonic plasma flow
  simulation for charged bodies such as dust particles and spacecraft,
  \textit{Phys. Rev. E}, \textit{52}, 5312.

\bibitem[{\textit{Miloch et~al.}(2007)\textit{Miloch, P{\'e}cseli, and
  Trulsen}}]{Miloch_Pecseli_Trulsen_2007}
Miloch, W.~J., H.~L. P{\'e}cseli, and J.~Trulsen (2007), Numerical simulations
  of the charging of dust particles by contact with hot plasmas,
  \textit{Nonlin. Processes Geophys.}, \textit{14}, 575--586.

\bibitem[{\textit{Miloch et~al.}(2008{\natexlab{a}})\textit{Miloch, Trulsen,
  and P{\'e}cseli}}]{Miloch_Pecseli_Trulsen_2008}
Miloch, W.~J., J.~Trulsen, and H.~L. P{\'e}cseli (2008{\natexlab{a}}),
  Numerical studies of ion focusing behind macroscopic obstacles in a
  supersonic plasma flow, \textit{Phys. Rev. E}, \textit{77}, 056408.

\bibitem[{\textit{Miloch et~al.}(2008{\natexlab{b}})\textit{Miloch, Vladimirov,
  P{\'e}cseli, and Trulsen}}]{Miloch_Vladimirov_2008}
Miloch, W.~J., S.~V. Vladimirov, H.~L. P{\'e}cseli, and J.~Trulsen
  (2008{\natexlab{b}}), Wake behind dust grains in flowing plasmas with a
  directed photon flux, \textit{Phys. Rev. E}, \textit{77}, 065401(R).

\bibitem[{\textit{Miloch et~al.}(2008{\natexlab{c}})\textit{Miloch, Vladimirov,
  P{\'e}cseli, and Trulsen}}]{Miloch_Vladimirov_2008b}
Miloch, W.~J., S.~V. Vladimirov, H.~L. P{\'e}cseli, and J.~Trulsen
  (2008{\natexlab{c}}), Numerical simulations of potential distribution for
  elongated, insulating dust being charged by drifting plasmas, \textit{Phys.
  Rev. E}, \textit{78}, 036411.

\bibitem[{\textit{Roussel and Berthelier}(2004)}]{Roussel_Berthelier_2004}
Roussel, J.~F., and J.~J. Berthelier (2004), A study of the electrical charging
  of the Rosetta orbiter: 2. numerical model, \textit{J. Geophys. Res.},
  \textit{109}, A01104.

\bibitem[{\textit{Shukla and Mamun}(2002)}]{Shukla_Mamun_2002}
Shukla, P.~K., and A.~A. Mamun (2002), \textit{Introduction to Dusty Plasmas},
  Institute of Physics Publishing, Bristol.

\bibitem[{\textit{Svenes and Tr{\o}im}(1994)}]{Svenes_Troim_1994}
Svenes, K.~R., and J.~Tr{\o}im (1994), Laboratory simulation of vehicle-plasma
  interaction in low Earth orbit, \textit{Planet. Space Sci.}, \textit{42},
  81--94.

\bibitem[{\textit{Tautz and Lai}(2006)}]{Tautz_Lai_2006}
Tautz, M., and S.~T. Lai (2006), Analytical models for a spherical satellite
  charging in sunlight at any spin rate, \textit{Ann. Geophys.}, \textit{24},
  2599--2610.

\bibitem[{\textit{Tautz and Lai}(2007)}]{Tautz_Lai_2007}
Tautz, M., and S.~T. Lai (2007), Charging of fast spinning spheroidal
  satellites in sunlight, \textit{J. Appl. Phys.}, \textit{102}, 024905.

\bibitem[{\textit{Vladimirov and Ishihara}(1996)}]{Vladimirov_Ishihara_1996}
Vladimirov, S.~V., and O.~Ishihara (1996), On plasma crystal formation,
  \textit{Physics of Plasmas}, \textit{3}(2), 444--446, \doi{10.1063/1.871895}.

\bibitem[{\textit{Vladimirov and Nambu}(1995)}]{Vladimirov_Nambu_1995}
Vladimirov, S.~V., and M.~Nambu (1995), Attraction of charged particulates in
  plasmas with finite flows, \textit{Phys. Rev. E}, \textit{52}(3), R2172.

\bibitem[{\textit{Vladimirov et~al.}(2005)\textit{Vladimirov, Ostrikov, and
  Samarian}}]{Vladimirov_Ostrikov_2005}
Vladimirov, S.~V., K.~Ostrikov, and A.~A. Samarian (2005), \textit{Physics and
  applications of complex plasmas}, Imperial College Press, London.

\end{thebibliography}

%
%
%
%
%
%
%
%


%
%

\end{article}
\newpage

%

\begin{table}[!htb]
\caption{The mean charge value on the spinning insulating object in the presence of photo\-emission for $\zeta=100$. The charge is presented in units of elementary two-dimensional charge: $q_0=e \sqrt{n}$, where $e$ is the elementary charge in a three dimensional system and $n$ is a plasma density in a two dimensional system. }
\label{tab:meanq}
\begin{displaymath}
\begin{array}{lccc}
\hline
\hline
 & {\omega=0.5\pi} & \omega=2.0\pi & \omega=3.0\pi \\
 v_d~(C_s) & q~(q_0) & q~(q_0)& q~(q_0)   \\
\hline
0 &   -247 &  11 &  5 \\
0.5 & -368   & 2 &  -7 \\
1.5 & -648 & -180 & -120 \\
 \hline
 \hline
\end{array}
\end{displaymath}
\end{table}




%
%
%
%
%
%


\end{document}